\RequirePackage[mathlines]{lineno} 
\documentclass[a4paper,11pt]{article}
\pdfoutput=1 

\usepackage{jheppub} 

\usepackage[T1]{fontenc} 
\usepackage{comment}
\usepackage{siunitx}
\usepackage{booktabs}
\usepackage{hyperref}
\hypersetup{hidelinks,
	colorlinks=true,
	allcolors=blue,
	pdfstartview=Fit,
	breaklinks=true}
	
\newcommand{\BESIIIorcid}[1]{%
  \href{https://orcid.org/#1}{%
    \hspace*{0.1em}%
    \raisebox{-0.45ex}{\includegraphics[width=1em]{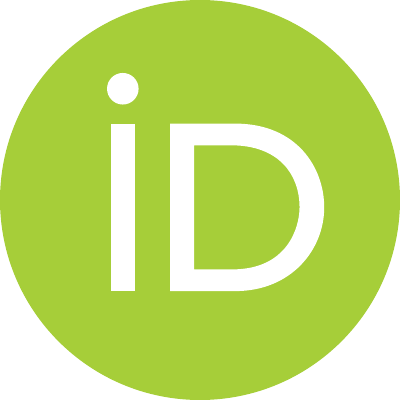}}%
  }%
}

\usepackage{threeparttable}
\usepackage{multirow}
\usepackage{subfigure}
\usepackage{float}
\let\oldequation\equation
\let\oldendequation\endequation
\renewenvironment{equation}
  {\linenomathNonumbers\oldequation}
  {\oldendequation\endlinenomath}

\def \ee   {e^+e^-}

\def \pim  {\pi^-}

\def \gevcc{~\mbox{GeV/$c^2$}}

\def \mevcc{~\mbox{MeV/$c^2$}}

\def\BR{\mathcal{B}}
\def\jpsi{J/\psi}
\def\Ks{K_S^0}

\def\ee{e^+e^-}

\def \ifb  {\mbox{fb$^{-1}$}}

\def\pippim{\pi^+\pi^-}

\def\BR{\mathcal{B}}
\def\jpsi{J/\psi}
\def\Ks{K_S^0}

\def\Lc{\Lambda_c^+}
\def\Lcm{\bar{\Lambda}_{c}^{-}}
\def \lcpetap{\Lambda_c^+\to p \eta'}
\def\lcpomg{\Lambda_c^+\to p \omega}
\def\etapdecay{\eta'\to \pi^+\pi^-\gamma}
\def\omgdecay{\omega\to \pi^+\pi^-\pi^0 }
\def\pigg{\pi^0
\to \gamma \gamma}

\def\mbc{M_{\rm{BC}}}
\def\dE{\Delta E}

\def\sumnominal{0.55\pm 0.22_{\rm{stat.}} \pm 0.05_{\rm{syst.}}}
\def\sumnominalnew{0.55\pm 0.22  \pm 0.05 }

\def\signifIni{ 3.4 \sigma}

\def\BFthiswork{ 6.15\pm 2.45_{\rm{stat.}} \pm 0.53_{\rm{syst.}} \pm 1.16 _{\rm{ref.}}}

\begin{document}

\title{\boldmath Measurement of the  singly Cabibbo-suppressed decay $\Lambda_c^+\to p\eta'$ with Deep Learning }

\collaborationImg{\includegraphics[height=30mm,angle=90]{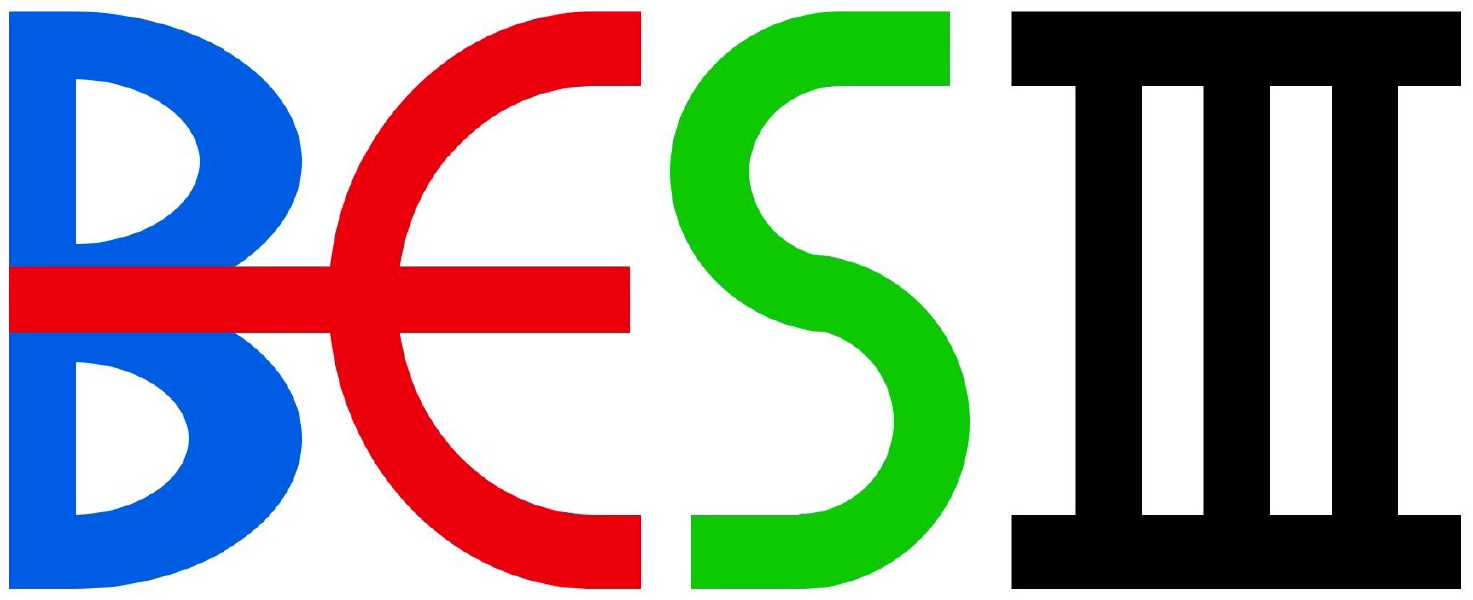}}
\collaboration{The BESIII collaboration}
\emailAdd{besiii-publications@ihep.ac.cn}

\abstract{
Using $4.5$ fb$^{-1}$ of $e^+e^-$ collision data collected with the BESIII detector at  center-of-mass energies from 4.600 to 4.699 GeV, we report a measurement of the singly Cabibbo-suppressed decay  $\Lambda_c^+ \to p\eta'$  with the single-tag method. To effectively distinguish the signal from the large backgrounds, we exploit a deep-learning classifier built on a Transformer-based neural network. Extensive validation and uncertainty quantification are carried out. The $\Lambda^+_c\to p\eta^\prime$ signal is observed with a statistical significance of $\signifIni$. The ratio of branching fractions of $\BR(\lcpetap)/\BR(\lcpomg)$= $\sumnominalnew$ is obtained, where the first uncertainty is statistical and the second systematic. 

}


\maketitle
\flushbottom

\section{Introduction}
\label{sec:introduction}
\hspace{1.5em} 
Weak decays of charmed baryons provide a unique laboratory for probing strong and weak dynamics in the charm sector~\cite{weak_decay,charm_review}. The corresponding decay amplitudes are usually split into factorizable and non-factorisable contributions~\cite{fac1,fac2}. In charmed-meson decays the latter are strongly suppressed and can often be neglected~\cite{suppress}, whereas in charmed-baryon decays W-exchange topologies give sizable non-factorisable contributions, complicating any theoretical calculation.

\begin{figure}[htbp] 
\centering
\subfigure[Internal  $W$-emission] {
\label{fig:fm_we}
\includegraphics[width=0.45\columnwidth]
{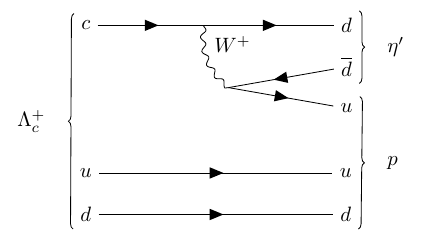}
}
\subfigure[ $W$-exchange] {
\label{fig:fm_wex}
\includegraphics[width=0.45\columnwidth]
{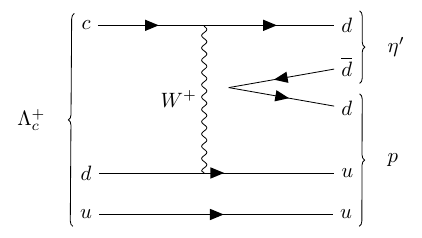}
}
\caption{Topological diagrams of the $\lcpetap$ decay.}
\label{fig:fmplot}
\end{figure}

 For the two-body
singly Cabibbo-suppressed (SCS) decay $\lcpetap$, the decay amplitude receives contributions from both factorizable $W$-emission diagram and non-factorizable $W$-exchange diagram, as shown in Fig.~\ref{fig:fmplot}. 
Various phenomenological approaches, including the constituent-quark model~\cite{constit}, the topological-diagram framework~\cite{su3_2020,topodiagram}, and SU(3)-flavour symmetry analyses~\cite{su31,su32,su3_2019,su3_2022,su3_2023,XHY}, predict the branching fraction of $\lcpetap$ in the range $\mathcal{O}(10^{-4})-\mathcal{O}(10^{-3})$, but the spread is large. The topological approach furthermore predicts $CP$ violation at the per-mille level~\cite{topodiagram}. Experimental information on the BF is therefore essential to refine these models.

Significant progress has been made in the measurement of the BF of $\lcpetap$.
The Belle Collaboration first observed this decay, reconstructing $\eta'\to\pi^+\pi^-\eta$ with a significance of 5.4$\sigma$ and measuring its relative BF~\cite{belle_petap}. The BESIII Collaboration  reported  evidence for this decay with a  significance of 3.6$\sigma$ using two dominant $\eta'$ decay modes~\cite{bes_petap}. In the $\eta'\to\pippim\gamma$ mode, the corresponding signal yield was $4.3^{+2.6}_{-2.2}$. In the $\eta'\to\pippim\eta$ mode, the signal yield was determined to be $4.9^{+3.2}_{-2.6}$, where the $\eta$ meson was treated as a missing particle rather than explicitly reconstructed  with individual final states  to enhance detection efficiency.
The main experimental challenge is the huge combinatorial background in $e^+e^-$ collisions. The previous BESIII analysis addressed this issue using a double-tag (DT) method~\cite{bes_petap,markiii}, in which the $\Lcm$ in the recoil side of $\Lc$ was fully reconstructed with exclusive decay modes to suppress backgrounds. 
Although this DT method is effective in background reduction, it suffers from low efficiency and limited statistics.
A single-tag (ST) approach,   in which the  $\lcpetap$ decay is reconstructed without imposing
constraints on the rest of the $\ee$ collision event,   greatly increases efficiency, but the background rises steeply. In this work we adopt the ST strategy and separate signal from background with a deep neural network (DNN) inspired by recent LHC jet-tagging applications~\cite{lhc1,lhc2} and already successfully used in other BESIII analyses~\cite{nenu,ppi0,genv}. In this analysis, the $\eta'$ meson is reconstructed exclusively via the $\eta' \to \pi^+\pi^-\gamma$ decay mode. Another   decay mode $\eta'\to\pippim\eta$ is not included, as its smaller branching fraction and lower selection efficiency result in a negligible expected signal yield under the current strategy.

In this work, we report a new measurement of the SCS 
decay $\lcpetap$ using 4.5 $\ifb$ of $\ee$ collision data
collected with the BESIII detector at seven center-of-mass
(c.m.) energies between 4.600 and 4.699 GeV~\cite{lum1,lum2}. 
The ST sample is classified with a DNN, and $\Lambda_c^+\to p\omega$ serves as the normalization channel. 
Using the ratio $\BR(\Lambda_c^+\to p\eta')/\BR(\Lambda_c^+\to p\omega)$ cancels many systematic uncertainties. Although the data set is identical to that of Ref.~\cite{bes_petap}, the analysis approach is independent. Charge-conjugate modes are implied throughout this paper.

\section{BESIII detector and Monte Carlo simulation}
\label{sec:detector}
\hspace{1.5em}
The BESIII detector~\cite{Ablikim:2009aa} records symmetric $e^+e^-$ collisions 
provided by the BEPCII storage ring~\cite{Yu:IPAC2016-TUYA01} in the c.m. energy ($\sqrt{s}$) range from 1.84 to 4.95~GeV, with a peak luminosity of $1.1\times10^{33}\;\text{cm}^{-2}\text{s}^{-1}$ achieved at $\sqrt{s} = 3.773\;\text{GeV}$.
BESIII has collected large data samples in this energy region~\cite{Ablikim:2019hff}. The cylindrical core of the BESIII detector covers 93\% of the full solid angle and consists of a helium-based
 multilayer drift chamber~(MDC), a plastic scintillator time-of-flight
system~(TOF), and a CsI(Tl) electromagnetic calorimeter~(EMC),
which are 
all enclosed in a superconducting solenoidal magnet providing a 1.0~T magnetic field.
The solenoid is supported by an
octagonal flux-return yoke with resistive plate counter muon
identification modules interleaved with steel. 

The charged-particle momentum resolution at $1~{\rm GeV}/c$ is
$0.5\%$, and the ionization energy loss (d$E$/d$x$) resolution is $6\%$ for electrons. The EMC measures photon energies with a
resolution of $2.5\%$ ($5\%$) at $1$~GeV in the barrel (end-cap)
region. The time resolution in the TOF barrel region is 68~ps, while
that in the end-cap region is 110~ps. 
The end-cap TOF system was upgraded in 2015 using multi-gap resistive plate chamber
technology, providing a time resolution of 60~ps, which benefits 87\% of the data used in this analysis~\cite{etof1, etof2}.

Monte Carlo (MC) simulated data samples produced with the {\sc geant4}-based~\cite{geant4} software package, which includes the geometric and material description of the BESIII detector~\cite{geo1,geo2} and the detector response, are used to determine detection efficiencies and estimate backgrounds. The simulation models the beam energy spread and initial state radiation (ISR) in the $e^+e^-$ annihilations with the generator {\sc kkmc}~\cite{ref:kkmc1, ref:kkmc2}. 
The inclusive MC sample includes the production of  the $\Lc\Lcm$ pairs, the open charmed processes, the ISR production of vector charmonium(-like) states, and the continuum QCD processes $\ee\to q\bar{q}$ ($q=u,d,s$) incorporated in {\sc kkmc}.
All particle decays are modeled with {\sc evtgen}~\cite{ref:evtgen1, ref:evtgen2} using the BFs either taken from the Particle Data Group~(PDG)~\cite{pdg:2022}, when available, or otherwise estimated with {\sc lundcharm}~\cite{ref:lundcharm1, ref:lundcharm2}. Final state radiation from charged final state particles is incorporated using the {\sc photos} package~\cite{photos}.
The signal MC samples for $\lcpetap$ and $\lcpomg$ are generated using phase space model. The subsequent decay of $\etapdecay$ is  simulated by a model that considers both the $\rho-\omega$ interference and the box
anomaly~\cite{geneta}, while the  decay of $\omgdecay$ is modeled with an $\omega$ Dalitz
model~\cite{genomg}.

\section{Analysis method}
\label{sec:analysis}
\hspace{1.5em} 
In this analysis, 
 the relative BF of $\lcpetap$ with respect to $\lcpomg$ is measured, where the  $\Lc$ baryon is produced in pairs through $\ee\to \Lc\Lcm$.    Since the data sets utilized in this work are taken at the
energy just above the $\Lc\Lcm$
mass threshold, the $\Lc\Lcm$
pairs are produced without any
accompanying hadrons. 
In order to obtain   higher detection efficiencies along with  higher yields,  the ST method is utilized. 
Under this strategy, the hadronic background would increase significantly due to the lack of additional constraint on the $\Lcm$ in the recoil side. The DNN is then invoked
for  a three-class classification of signal, $\Lc\Lcm$  backgrounds and   hadronic backgrounds. The $\eta'$ candidate is selected  via its dominant decay mode $\eta'\to\pippim\gamma$.  

The normalization channel $\Lambda_c^+\to p\omega$, $\omega\to\pi^+\pi^-\pi^0$ is selected analogously. To reduce the systematic uncertainties introduced by the DNN method, the  relative BF of $\lcpetap$ with respect to $\lcpomg$ is determined by
\begin{equation}
\frac{\mathcal{B}(\lcpetap)}{\mathcal{B}(\lcpomg)}=\frac{\mathcal{B}_{\omgdecay}\cdot \BR_{\pigg}}{\mathcal{B}_{\etapdecay}} \cdot \frac{\sum_i\left(N_{\mathrm{sig}}^i / \varepsilon_{\mathrm{sig}}^i\right)_{\lcpetap}}{\sum_i\left(N_{\mathrm{sig}}^i / \varepsilon_{\mathrm{sig}}^i\right)_{\lcpomg}},
\label{eq:relative_ratio}
\end{equation}
where the superscript $i$ denotes the c.m. energy,  $N^i_{\rm{sig}}$ stands for  the signal yields,  $\varepsilon^i_{\rm{sig}}$ 
 is the signal efficiency as determined by MC simulation, and $\BR_{\etapdecay}$, $\BR_{\omgdecay}$ and $\BR_{\pi^0\to\gamma\gamma}$ are the
BFs of the $\etapdecay$, $\omgdecay$ and $\pigg$ decays taken from the PDG~\cite{pdg:2022},
respectively. 


\section{Initial event selection}
\label{sec:evt_select}
\hspace{1.5em} 
To select the candidates for  $\lcpetap$, with $\etapdecay$, we reconstruct the events from combinations of charged tracks
and photon candidates  that satisfy the following selection criteria.

Charged tracks detected   in the MDC are required to be within a polar angle ($\theta$) range of $|\rm{cos\theta}|<0.93$, where $\theta$ is defined with respect to the symmetry axis of the MDC ($z$ axis). For charged tracks, the distance of the closest approach to the  interaction point (IP) 
must be less than 10\,cm along the $z$ axis  and less than 1\,cm in the transverse plane.  Particle identification~(PID) for charged tracks is implemented by combining measurements of the d$E$/d$x$  in the MDC and the flight time in the TOF to form likelihoods $\mathcal{L}(h)~(h=p, K,$ or $\pi)$ for each hadron $h$ hypothesis.
Charged tracks are 
 identified as protons, when the proton hypothesis has the largest likelihood, i.e.,  $\mathcal{L}(p)>\mathcal{L}(\pi)$,  $\mathcal{L}(p)>\mathcal{L}(K)$, while pion candidates are  identified by requiring  $\mathcal{L}(\pi)>\mathcal{L}(K)$.

Photon candidates from $\eta'$  decays are reconstructed using electromagnetic showers
produced in the EMC. The deposited energy of each shower is required to be greater than
25 MeV in the barrel region ($|\cos\theta|$ < 0.80) and 50 MeV in the end-cap region (0.86 <
$|\cos\theta|$  < 0.92). In order to suppress the  electronic noise and showers unrelated to the event, the difference between the EMC time and event start
time~\cite{EST} is required to be within 700 ns. To exclude
showers that originate from charged tracks, the angles subtended by the EMC shower and the position
of the closest charged track at the EMC must be
greater than $10^{\circ}$ as measured from the IP.

To reconstruct the  $\eta'$ meson, the invariant mass of $\pippim\gamma$, $M(\pippim\gamma)$, is required to be in the range
of [0.90, 1.00]$\gevcc$.
In addition, to suppress the backgrounds from  the intermediate decays $\Lambda\to p \pi^-$ and $K_S^0 \to \pippim$,  the  invariant masses of $p\pim$ and $\pippim$  are required to lie  outside the regions of  [1.10, 1.15]$\gevcc$   and [0.48, 0.51]$\gevcc$, respectively.

To further identify signal candidates, we use the beam energy constrained mass $\mbc$ and the energy
difference $\dE$, defined as:
\begin{equation}
\begin{aligned}
\Delta E & =E_{\Lambda_c^{+}}-E_{\text {beam }}, \\
M_{\mathrm{BC}} & =\sqrt{E_{\text {beam }}^2 / c^4-\left|\vec{p}_{\Lambda_c^{+}}\right|^2 / c^2,} 
\end{aligned}
\label{eq:mbcdE}
\end{equation}
where $E_{\text {beam }}$ is the beam energy, and $\vec{p}_{\Lambda_c^{+}}$ and $E_{\Lambda_c^{+}}$ are the momentum and energy of $\Lc$ candidate in the $\ee$ rest frame, respectively. 
For a correctly reconstructed $\Lc$
candidate, one expects $\mbc$ to be around the nominal $\Lc$ mass~\cite{pdg:2022} and $\dE$
to be around zero. 
The combination with the minimum $|\dE|$ is retained if multiple combinations exist in an event.
The $\dE$ of the event candidate is further required to satisfy $-65$ MeV < $\dE$ < $60$ MeV.

The candidates for the reference sample of $\lcpomg$  are selected similarly, including the selections of  pion, proton, and photon candidates. For $\pi^0$ candidates, the invariant mass of the photon pair ($M_{\gamma\gamma}$) is required to be in the range of [0.115, 0.150]$\gevcc$. To improve the energy resolution, a one-constraint  kinematic fit is utilized to constrain the $M_{\gamma\gamma}$ to the nominal $\pi^0$ mass~\cite{pdg:2022}. The updated four-momentum by the kinematic fit
 will be used in the further analysis.
The $\omega$ candidates are reconstructed through the decay of $\omgdecay$, and the invariant mass of $\pi^+\pi^-\pi^0$, $M(\pi^+\pi^-\pi^0)$,  is required to be within [0.73, 0.83]$\gevcc$. Possible backgrounds from $\Lambda$ and $\Ks$ are rejected using the same requirements as in the signal selection criteria. 
In addition, the backgrounds from $\Sigma^+ \rightarrow p \pi^0$ are vetoed by excluding the $M(p\pi^0)$ mass region  of [1.17, 1.20]$\gevcc$. 
If multiple combinations  are present in an event, the one with the smallest $|\dE|$ is retained for further analysis, and its $\dE$  is required to satisfy $-40$ MeV < $\dE$ < $30$ MeV.

After applying the above selections, the $\mbc$ distributions for both data and inclusive MC samples for the signal and reference processes
 are shown in Fig.~\ref{fig:ST_bkg}.
According to the plots, the MC simulation describes the data well.
Analysis of the inclusive MC sample indicates that the primary backgrounds are categorized into two parts:  $\Lc\Lcm$ non-signal background and hadronic  background, in which the latter one is dominant. The $\Lc\Lcm$ non-signal background originates from the $\Lc$ baryon decays, while the hadronic background originates  from  
$\ee\to q\bar{q} $ $(q=u, d, s)$.
For  $\lcpomg$, the backgrounds are categorized into three components: $\Lc\Lcm$ non-$\omega$ backgrounds, $\Lc\Lcm$ other
backgrounds and hadronic  backgrounds, in which the last one is dominant.  The $\Lc\Lcm$ non-$\omega$  background includes the peaking contribution from the $\Lc$ decays without an $\omega$ meson in the final state and is studied using the $\omega$ sideband region, defined as  (0.62, 0.72)$\gevcc$ $\cup$ (0.85, 0.90)$\gevcc$  in the $M(\pi^+\pi^-\pi^0)$ distribution. 
Given the current prevailing background level, we are unable  to extract the  signal yields  for $\lcpetap$ from the $\mbc$ distribution yet, and thereby require 
the DNN classifier as elaborated
below.

\begin{figure}[htbp] 
\centering
\subfigure[] {
\label{fig:petap_st}
\includegraphics[width=0.45\columnwidth]
{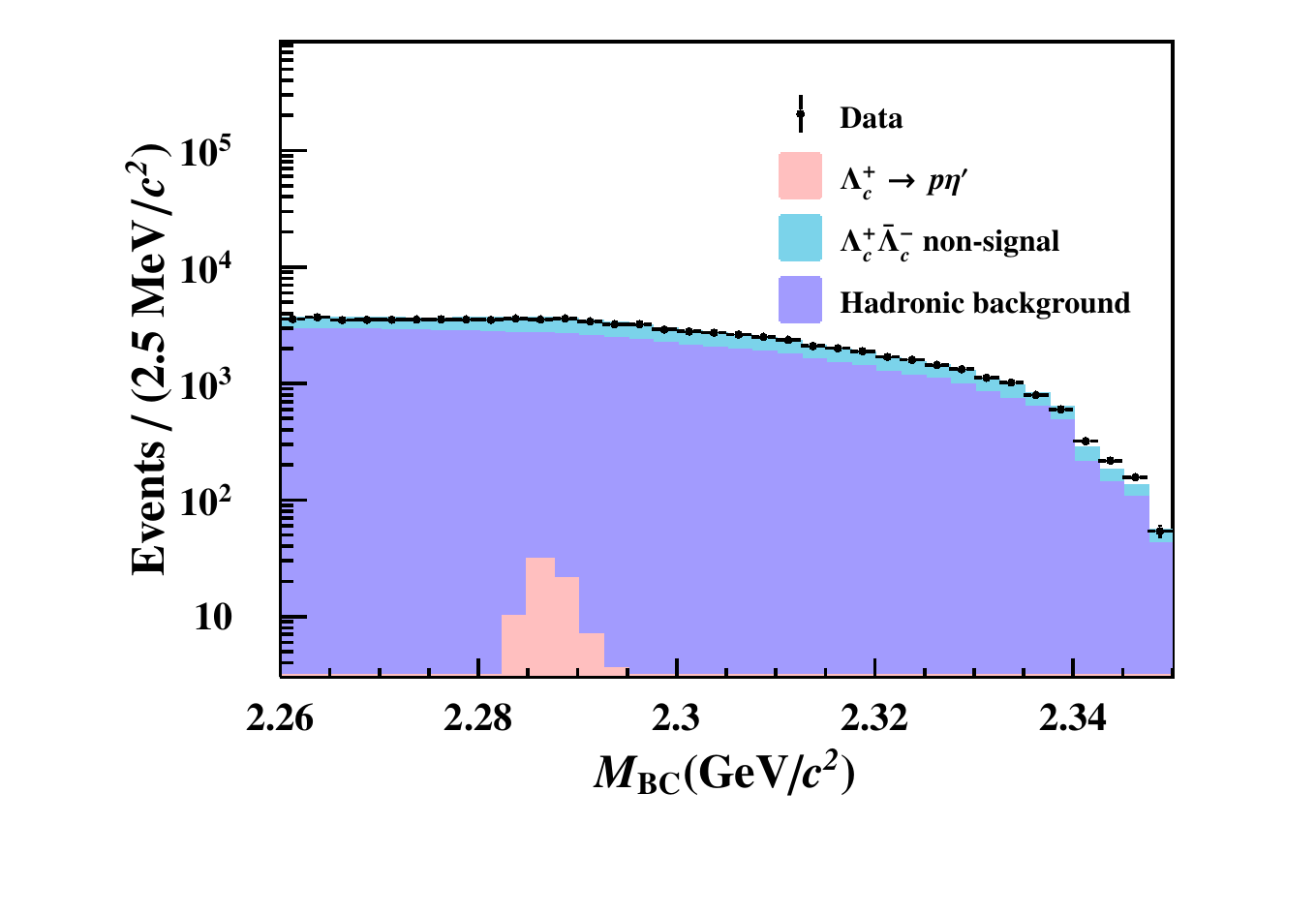}
}
\subfigure[] {
\label{fig:pomg_st}
\includegraphics[width=0.45\columnwidth]
{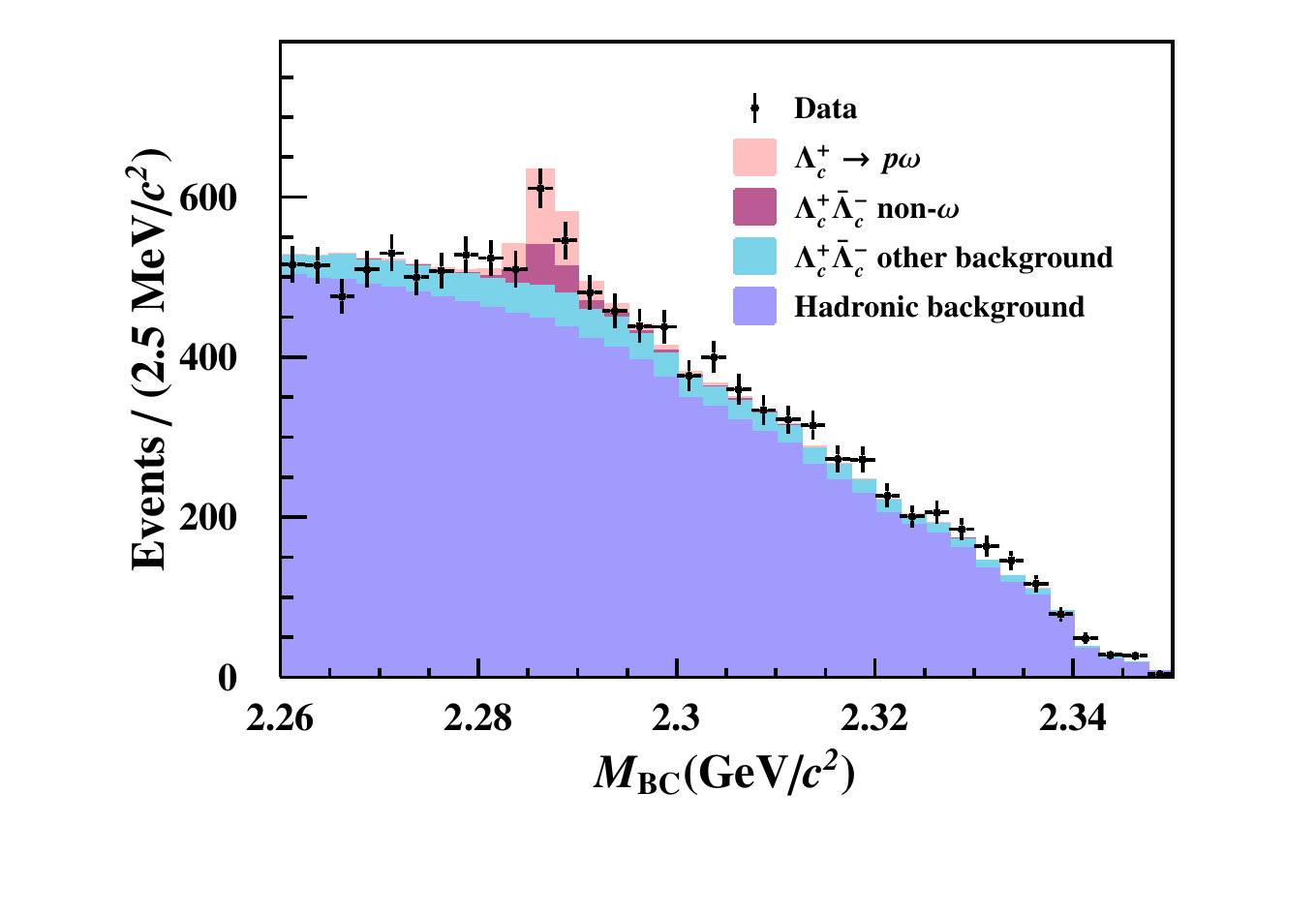}
}
\caption{Distributions of $\mbc$ for (a) $\lcpetap$ and
(b) $\lcpomg$ after preselection before implementing deep learning. The black
points with error bars represent data and  the pink shaded histograms  represent the signal
components. The purple shaded histograms denote the hadronic backgrounds, while the blue
shaded histograms correspond to the $\Lambda_c^+\bar{\Lambda}_c^-$ non-signal backgrounds. In (b), the magenta shaded histograms denote the $\Lambda_c^+\bar{\Lambda}_c^-$ non-$\omega$ backgrounds, while the blue shaded histograms  show the remaining   $\Lc\Lcm$ contributions. }
\label{fig:ST_bkg}
\end{figure}

\vspace{-0.0cm}

\section{Deep-learning-based signal identification}
\label{sec:dnn}
\hspace{1.5em} 
Although the event selection criteria derived from the ST method significantly improve detection efficiency compared to a DT analysis, they also introduce substantial contamination from hadronic backgrounds.
Traditional cut-based approaches  depend on manually constructed variables rooted in empirical understanding of the underlying physics, and are  limited  in effectively distinguishing signal from background. Motivated by the remarkable success of deep learning techniques in jet tagging tasks~\cite{lhc1,lhc2}, we explore the application of a DNN-based classification method for this analysis. The strategy of this work builds upon the analysis framework established in Ref.~\cite{ppi0},  and is specifically tailored to accommodate the distinct signal mode and the more complex background components in this study.

In this work, a DNN is  trained for a three-class classification of  signal,  $\Lc\Lcm$ non-signal background and hadronic background. Specifically, for the $\lcpomg$ decay, the $\Lc\Lcm$ non-$\omega$ background and $\Lc\Lcm$ other backgrounds are  merged as  a single  class:  the $\Lc\Lcm$ non-signal background.
The training sample is assembled from dedicated MC simulations. The events
are categorized into three groups with equal statistics:  signal,  $\Lc\Lcm$ non-signal background, and hadronic background. Events are generated at different c.m.\ energies in proportion to their respective yields observed in the data.  For the $\Lambda_c^+ \to p\eta'$ decay, the total training sample contains approximately $1.5 \times 10^6$ events, while for $\Lambda_c^+ \to p\omega$, around $1.0 \times 10^6$ events are used. 
80\% of the events in the samples are used for training and the remaining 20\% are reserved for independent validation.
It should be noted that  a combined dataset including both the  $\lcpetap$ and 
$\lcpomg$ candidates  is fed into one unified DNN. This strategy ensures that the model achieves a generalization  performance across both decay modes, which is essential for employing the $\Lambda_c^+ \to p\omega$ candidates  as  reference.

The input information to the DNN 
consists of all charged
tracks reconstructed in the MDC and isolated showers
clustered in the EMC. These particles are represented 
as a point cloud~\cite{pointcloud} structure, i.e., an unordered
and variable-sized set of the points in a high dimensional
feature space. Track features include polar and azimuthal angles, charge, momentum and helical-track parameters. Low-level quantities from MDC, TOF and EMC are included to encode particle-identification and reconstruction quality. Shower features comprise angles, energy, number of fired crystals, shower time and lateral moments. Agreement between data and MC is verified for every input~\cite{ppi0}. 

The architecture of the DNN is primarily based on 
the Particle Transformer (ParT)~\cite{pT}. The model
hyperparameters are optimized to maximize the DNN
performance while avoiding overfitting. To further enhance the model’s robustness and generalization capability, a machine learning technique called {\it{model
ensemble}} is employed.
 With some
randomness factors incorporated in the training such as
network initialization, batch processing sequence, and
dropout~\cite{dropout} mechanism, ten independent DNNs are trained in parallel. The outputs of these DNNs
are averaged for each event during inference, forming
an ensemble DNN that demonstrates superior robustness and
generalization ability compared to any single DNN model.

The DNN exports three scores between
[0, 1] to each event, reflecting the probabilities of the  event belonging to  the signal,  $\Lc\Lcm$ non-signal background, and hadronic background. Since the three scores always sum to one, only two of
them are independent. As the final event selection, we define a composite discriminator variable as $\mathcal{S}=$ ${\rm{score}_{sig}}\cdot (\rm{1-score_{\rm{hadronic}}})$, which combines the scores of the signal and hadronic background categories. Events with $\mathcal{S} > 0.98$ are retained.
This
value is optimized by maximizing the figure of merit $\frac{S}{\sqrt{S+B}}\times \frac{S}{S+B}$~\cite{ppi0}, where $S$ ($B$) denotes the expected signal (background) yields, as estimated from the inclusive MC sample and normalized to the full data samples.

After applying the deep learning discriminator $\mathcal{S}$, the distributions of $\mbc$ for both $\lcpetap$ and $\lcpomg$ are shown in Fig.~\ref{fig:fit_gnn}. 
The backgrounds are significantly reduced by more than two orders of magnitude in both channels while retaining approximately 40\% of the signal, greatly enhancing the sensitivity to the signal and allowing for a clear extraction of the  $\lcpetap$ signal.

\section{Signal extraction}
\label{sec:bkgana}
\hspace{1.5em} 
A simultaneous binned maximum-likelihood fit to the $M_{\rm BC}$ distributions of the two channels and all seven energy points is performed. The signal shape is taken from MC and convolved with a Gaussian whose parameters are left free.  For $\lcpetap$, the $\Lc\Lcm$  and hadronic backgrounds are modeled  using their respective MC-simulated shapes. The  $\Lambda^+_c\bar \Lambda^-_c$ background contribution is fixed based on MC studies, while  that of the hadronic background is left free. For $\lcpomg$, 
we follow the same procedure as described in Ref.~\cite{bes_pomg}, where the $\mbc$ distributions of the $\Lc$ decaying to non-$\omega$ processes are modeled using the shape derived from $\Lc\to p\pippim\pi^0$ MC samples, with yields  estimated using the   sideband region of  $M(\pippim\pi^0)$ in data.
The remaining $\Lc\Lcm$ background   and hadronic backgrounds are modeled using shapes derived from  dedicated MC simulation. The yields of the remaining $\Lc\Lcm$ background are fixed based on MC studies, while those from the remaining hadronic background float in the fit.
 According to Eq.~\ref{eq:relative_ratio}, the relative BF ratio $\BR(\lcpetap)/\BR(\lcpomg)$ is common among the fits for different c.m. energy points.  The signal efficiencies for each energy point are estimated with MC simulation, and summarized in Table~\ref{tab:eff_info}.
The fit yields $\BR(\lcpetap)/\BR(\lcpomg)$ = 0.55 $\pm$ 0.22, corresponding to total signal yields of 31.9 $\pm$ 14.3 for $\lcpetap$ and 69.8 $\pm$ 14.2 for $\lcpomg$.  Table~\ref{tab:sumResults} summarizes the fit results.  The statistical significance of $\lcpetap$ is determined to be 3.4$\sigma$, based on the change in likelihood and degrees of freedom in the
fits with and without the signal component. Figure~\ref{fig:fit_gnn} shows the fit projections combining all c.m. energies.

\begin{figure}[htbp!] 
\centering
\subfigure[] {
\label{fig:petap_fit}
\includegraphics[width=0.45\columnwidth]
{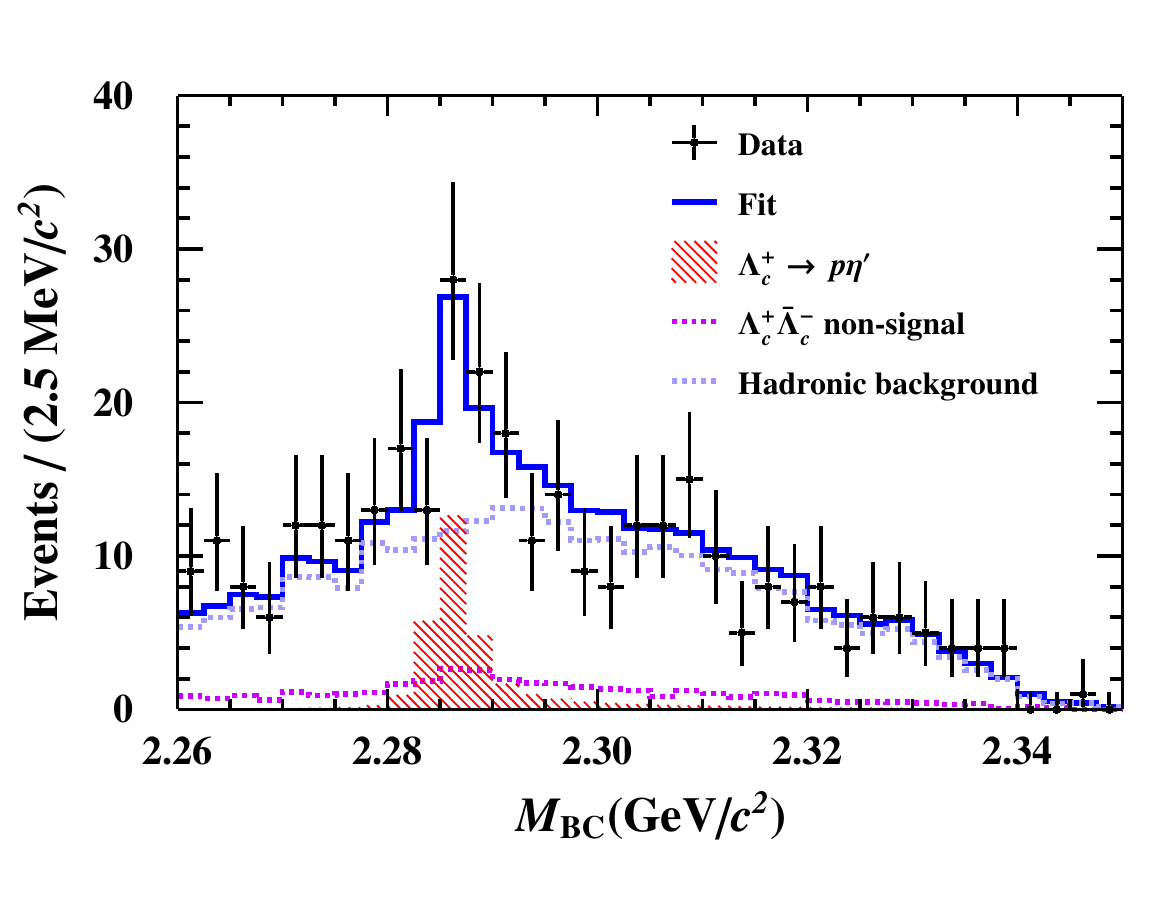}
}
\subfigure[] {
 \label{fig:pomg_fit}
\includegraphics[width=0.45\columnwidth]
{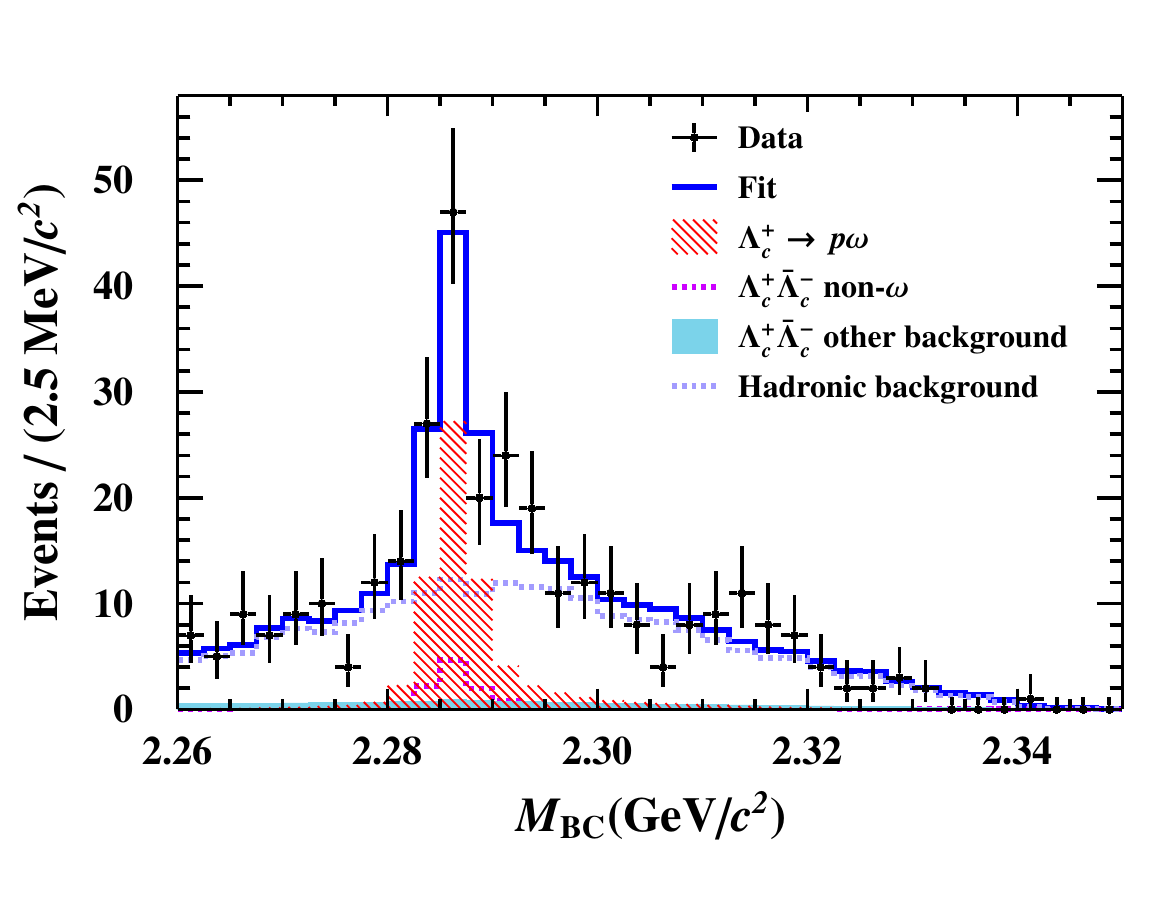}
}
\caption{Simultaneous fit results for (a) $\lcpetap$ and (b) $\lcpomg$ after the DNN-classifier. The points with error bars
denote data, the blue solid lines represent the total fit results, and the  purple dashed lines represent the hadronic background components. The red dashed areas represent the signal components.
 In (a),  the magenta dashed line represents the $\Lc\Lcm$ background components. In (b), the magenta dashed line denotes the  non-$\omega$ component, and the blue region represents the residual  backgrounds from $\Lc\Lcm$.
}
\label{fig:fit_gnn}
\end{figure}

\begin{table}[htbp]
\centering
\sisetup{separate-uncertainty = true, table-align-uncertainty = true}
\begin{tabular}{
    c
    S[table-format=2.2(2)]
    S[table-format=2.2(2)]
}
\toprule
$\sqrt{s}$ (MeV) &
{$\varepsilon_\text{sig}^i$ for $\Lambda_c^+ \to p\eta'$ (\%)} &
{$\varepsilon_\text{sig}^i$ for $\Lambda_c^+ \to p\omega$ (\%)} \\
\midrule
4600 & 24.27 \pm 0.07 & 10.54 \pm 0.03 \\
4612 & 20.73 \pm 0.04 &  7.70 \pm 0.03 \\
4626 & 20.58 \pm 0.04 &  7.69 \pm 0.03 \\
4640 & 19.53 \pm 0.04 &  7.72 \pm 0.03 \\
4660 & 18.87 \pm 0.04 &  7.34 \pm 0.03 \\
4680 & 16.85 \pm 0.04 &  6.68 \pm 0.02 \\
4700 & 15.22 \pm 0.04 &  5.99 \pm 0.02 \\
\bottomrule
\end{tabular}
\caption{The signal efficiencies for each c.m.~energy after implementing the DNN, where the uncertainties are statistical only.}
\label{tab:eff_info}
\end{table}

\begin{table}[htbp]
    \centering
    \begin{tabular}{c|c}
    \hline \hline
       Item  & Value \\
       \hline 
     $\BR(\lcpetap)/\BR(\lcpomg)$    & 0.55 $\pm$ 0.22  \\
   $N_{\lcpetap}$  & 31.9 $\pm$ 14.3  \\
 $N_{\lcpomg}$    &  69.8 $\pm$ 14.2  \\
     \hline \hline
    \end{tabular}
    \caption{The simultaneous fit results.  $\BR(\lcpetap)/\BR(\lcpomg)$ denotes the relative BF ratio between the $\lcpetap$  and $\lcpomg$ decays, $N_{\lcpetap}$ and $N_{\lcpomg}$ represent the total  signal yields for the $\lcpetap$ and $\lcpomg$ decays, respectively, where the uncertainties are statistical only.  }
    \label{tab:sumResults}
\end{table}

\section{Systematic uncertainties}
\label{sec:systematic}
\hspace{1.5em}
The strategy of this analysis effectively cancels   many potential systematic uncertainties. Specifically, the uncertainties associated with the total number of $\Lc\Lcm$ pairs are canceled, and  those from the
 $\dE$ requirements and the vetoes of  intermediate resonance decays (e.g., $\Sigma,\Ks$ and $\Lambda$) are significantly reduced in the relative BF according to Eq.~\ref{eq:relative_ratio}. The remaining
systematic uncertainties on the relative ratio, as summarized in Table~\ref{tab:sys}, are described below.
\begin{table}[htbp!]
    \centering
   \begin{tabular}{cc}
\hline \hline 
Source& Uncertainty~(in percent) \\  \hline 
 Tracking & 0.4 \\
 PID &  0.4\\
Reconstruction of $\pi^0$ & 2.5 \\ 
Reconstruction of photon & 0.5 \\
Resonances' mass window & 0.4 \\
MC model & 0.8 \\
Non-$\omega$ contribution & 2.5 \\
Neural network  & 4.6 \\
 Simultaneous fit &   6.2  \\
  Input BFs & 1.6 \\
 \hline
 Total & 8.7 \\
\hline \hline
\end{tabular}
    \caption{Relative systematic uncertainties in the BF measurement. }
    \label{tab:sys}
\end{table}

\begin{itemize}
    \item \emph{Tracking and PID}. The systematic uncertainties related to the tracking and PID for charged pions and proton are studied using a control sample of $J/\psi\to p \bar{p}\pi^+\pi^-$. The  MC samples are reweighted in two dimensions of transverse momentum  $P_t$ (momentum $P$)  and polar angle $\cos\theta$  to match the data of the control sample.
The remaining relative efficiency differences between the nominal and reweighted MC samples are assigned as the systematic uncertainties, which are 0.2\% (0.1\%) for proton (pion) tracking and 0.2\% (0.1\%) for proton (pion) PID. The total systematic
uncertainty from tracking (PID) is 0.4\%, with  individual uncertainties added linearly.
    \item \emph{$\pi^0$ reconstruction.}  The uncertainty from the $\pi^0$ reconstruction is evaluated to be 2.5\%  using a control sample of $D^0 \to K^- \pi^+ \pi^0$.
    
    \item \emph{Photon reconstruction}. The  systematic uncertainty  of the photon reconstruction in the EMC is evaluated using a control sample of $\jpsi\to\gamma\mu^+\mu^-$~\cite{showersys}, which is determined to be 0.5\%.
    \item \emph{$\omega$ and $\eta'$ mass window requirements}. To estimate the 
    systematic uncertainties related to the mass window requirements for $\omega$ and $\eta'$, the invariant mass distributions of $\pi^+ \pi^- \pi^0$ and $\pi^+ \pi^- \gamma$ in the MC samples are smeared with a Gaussian function with both mean and width set to 0.5$\mevcc$. The resulting variations on the
final results, 0.2\% for $\lcpetap$ and 0.4\% for $\lcpomg$, are taken as the  systematic uncertainties.
    \item \emph{MC model}. The  systematic uncertainty from the MC modeling is estimated by varying the decay polarization parameters of $\Lambda_c^+$ within their physically allowed ranges in the joint angular distributions for $\Lambda_c^+ \to p \eta'$ and $\Lambda_c^+ \to p \omega$. The largest deviations in efficiencies are 0.5\% for $\lcpomg$ and 0.6\% for $\lcpetap$.
    \item \emph{Non-$\omega$ contribution}. To evaluate  
    the systematic uncertainty  related to the estimation of non-$\omega$ background, alternative sideband regions are taken into consideration. Specifically, the nominal sideband regions are  varied by shifting each boundary by $\pm$10~MeV as alternatives. The largest change in the relative  BF ratio, 2.5\%, is assigned as the
systematic uncertainty.

    \item \emph{DNN classifier}.  The uncertainties related to the DNN output are evaluated following the method in Ref.~\cite{ppi0}, which considers two major sources  {\it{model uncertainty}} and  {\it{domain shift}}.  
The uncertainty due to model   arises from limitations within the model itself. In this work, the influence is estimated via the model ensemble technique, where the  relative shift in signal efficiency across the DNN classifier $\mathcal{S}$
during the ensemble models are derived. The standard deviation of these efficiencies, divided
by their average, 1.4\%, is assigned as the systematic uncertainty. Domain shift refers to the discrepancy between the data distribution used for training and inference,  reflecting potential data-MC inconsistencies. 
The leading  bias of such uncertainty  can be mostly canceled in the relative BF due to the similar final states of $\lcpetap$ and $\lcpomg$. This assumption has been validated using the control samples of $\Lambda_c^+\to  p K_S^0 \pi^0 $ and $\Lambda_c^+\to p K_S^0 \eta$, 
whose relative BF is found to be stable under any selection
on their DNN outputs. The largest  difference of resulted relative BF before and after implementing
DNN, 4.6\%, is assigned as the  systematic uncertainty.

    \item \emph{Fit model}. The systematic uncertainty associated with  the $\mbc$ fit is mainly due to the modeling of signal and background components. A bootstrap resampling method~\cite{boots1,boots2} is employed to quantify  the   uncertainty, yielding an uncertainty  of 6.2\%. The uncertainty of the hadronic background shape   is evaluated by replacing the MC-simulated shape with  an ARGUS function~\cite{argus}, whose parameters are fixed to the values obtained from fits to the corresponding distributions of inclusive MC samples.  The  uncertainty due to the  $\Lc\Lcm$ background
contribution is found to be  negligible.

    \item \emph{Input $\BR_{\rm{inter}}$}. The uncertainties of the  input BFs of  the intermediate decays $\etapdecay$, $\omgdecay$ and $\pi^0\to\gamma\gamma$ are taken from the PDG~\cite{pdg:2022}, which  are 1.4\%, 0.8\% and 0.1\%,  respectively.
\end{itemize}

Assuming that all sources are independent with each other, the total systematic uncertainty is determined to be 8.7\% by adding the individual uncertainties 
in quadrature, as summarized in Table~\ref{tab:sys}.

\begin{figure}[htbp!] \centering
	\setlength{\abovecaptionskip}{-1pt}
	\setlength{\belowcaptionskip}{10pt}
	\includegraphics[width=10.0cm]{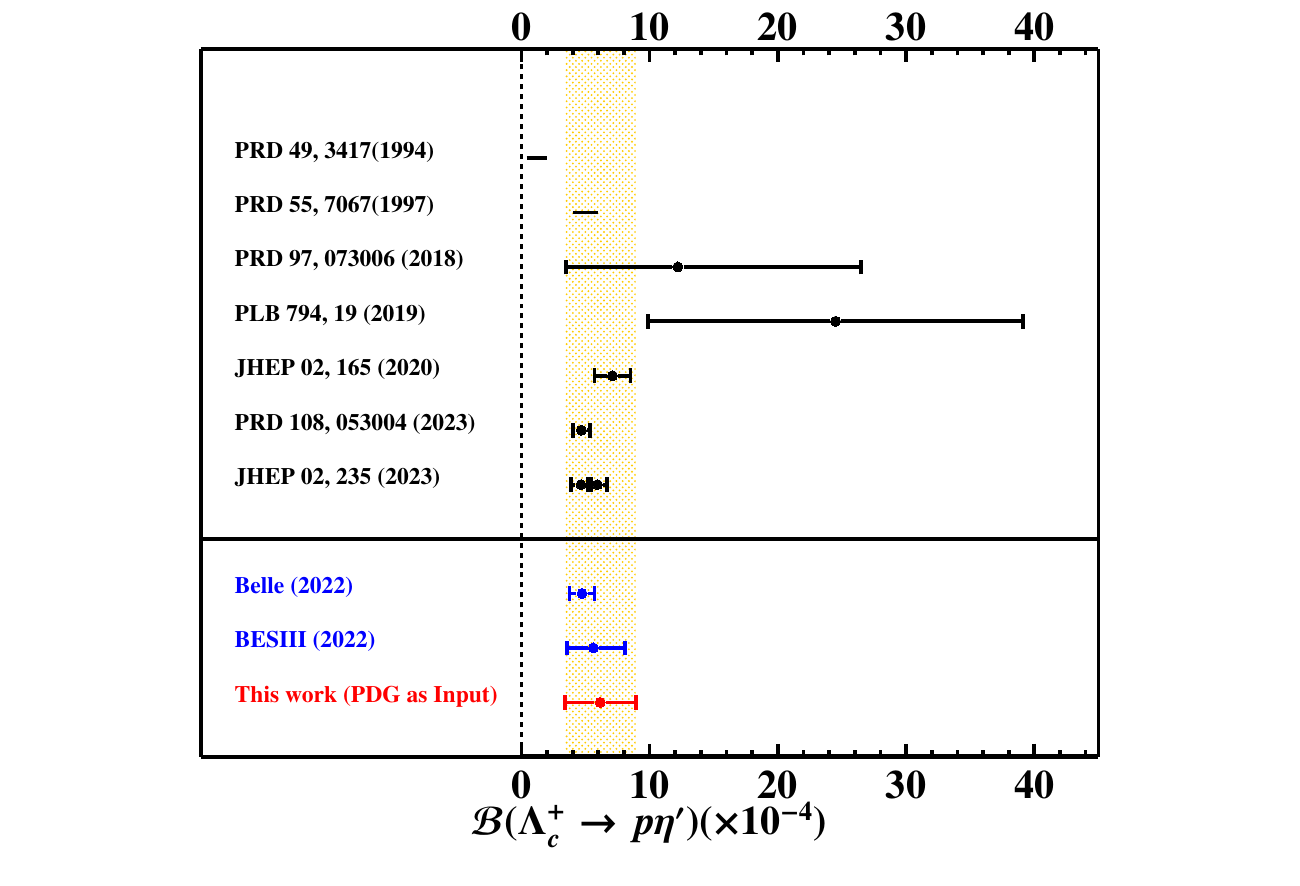}
	\caption{Comparison of our $\BR(\lcpetap)$ result~(red point with error bar, obtained using the world average $\BR(\lcpomg)$ from the PDG as input)   with  the previous
theoretical predictions (black points with error bars) and experimental measurements
(blue points with error bars) in time order. Note that the previous BESIII results were obtained using a DT method.  }
	\label{fig:compare}
\end{figure}

\section{Summary}
\label{sec:summary}
\hspace{1.5em}
Using 4.5 fb$^{-1}$ of $e^+e^-$ data at $\sqrt{s}=4.600$–$4.699$ GeV, BESIII has measured the SCS decay $\Lambda_c^+\to p\eta'$ with the ST method and a Transformer-based deep-learning classifier. The signal significance is   $\signifIni$ and the relative BF ratio between $\lcpetap$ and $\lcpomg$ is measured to be:
$\BR(\lcpetap)/\BR(\lcpomg)=\sumnominal$. 
With the world-average   $\BR(\lcpomg)$ = $(1.11\pm 0.21)\times 10^{-3}$ from the PDG~\cite{pdg:2022}, the BF of $\lcpetap$ is determined as $\BR(\lcpetap)=(\BFthiswork)\times 10^{-4}$. Figure~\ref{fig:compare} presents a comparison between the measured $\BR(\lcpetap)$ in this work, previous experimental results, and theoretical predictions.   Our ST measurement is consistent with the previous measurements~\cite{belle_petap,bes_petap} within 1$\sigma$. The ST approach, combined with the DNN suppressing background by two orders of magnitude, yields comparable precision to the earlier DT analysis with significantly higher signal yield. 
The measurement favors SU(3)-flavour and topological-model calculations~\cite{su3_2020,su31,su32,su3_2023,XHY} over the constituent-quark prediction~\cite{constit} and provides new input to refine theoretical descriptions of charmed-baryon decays.

\acknowledgments
\hspace{1.5em}
The BESIII Collaboration thanks the staff of BEPCII (https://cstr.cn/31109.02.BEPC) and the IHEP computing center for their strong support. This work is supported in part by National Key R\&D Program of China under Contracts Nos. 2023YFA1606000, 2023YFA1606704; National Natural Science Foundation of China (NSFC) under Contracts Nos. 11635010, 11935015, 11935016, 11935018, 12025502, 12035009, 12035013, 12061131003, 12192260, 12192261, 12192262, 12192263, 12192264, 12192265, 12221005, 12225509, 12235017, 12361141819; the Chinese Academy of Sciences (CAS) Large-Scale Scientific Facility Program; the Strategic Priority Research Program of Chinese Academy of Sciences under Contract No. XDA0480600; CAS under Contract No. YSBR-101; 100 Talents Program of CAS; The Institute of Nuclear and Particle Physics (INPAC) and Shanghai Key Laboratory for Particle Physics and Cosmology; ERC under Contract No. 758462; German Research Foundation DFG under Contract No. FOR5327; Istituto Nazionale di Fisica Nucleare, Italy; Knut and Alice Wallenberg Foundation under Contracts Nos. 2021.0174, 2021.0299; Ministry of Development of Turkey under Contract No. DPT2006K-120470; National Research Foundation of Korea under Contract No. NRF-2022R1A2C1092335; National Science and Technology fund of Mongolia; Polish National Science Centre under Contract No. 2024/53/B/ST2/00975; STFC (United Kingdom); Swedish Research Council under Contract No. 2019.04595; U. S. Department of Energy under Contract No. DE-FG02-05ER41374

\newpage
M.~Ablikim$^{1}$\BESIIIorcid{0000-0002-3935-619X},
M.~N.~Achasov$^{4,b}$\BESIIIorcid{0000-0002-9400-8622},
P.~Adlarson$^{81}$\BESIIIorcid{0000-0001-6280-3851},
X.~C.~Ai$^{86}$\BESIIIorcid{0000-0003-3856-2415},
R.~Aliberti$^{39}$\BESIIIorcid{0000-0003-3500-4012},
A.~Amoroso$^{80A,80C}$\BESIIIorcid{0000-0002-3095-8610},
Q.~An$^{77,64,\dagger}$,
Y.~Bai$^{62}$\BESIIIorcid{0000-0001-6593-5665},
O.~Bakina$^{40}$\BESIIIorcid{0009-0005-0719-7461},
Y.~Ban$^{50,g}$\BESIIIorcid{0000-0002-1912-0374}
% ...
\\
\vspace{0.2cm}
(BESIII Collaboration)\\
\vspace{0.2cm} {\it
$^{1}$ Institute of High Energy Physics, Beijing 100049, People's Republic of China\\
% ...
$^{4}$ Budker Institute of Nuclear Physics SB RAS (BINP), Novosibirsk 630090, Russia\\
% ...
$^{39}$ Johannes Gutenberg University of Mainz, Johann-Joachim-Becher-Weg 45, D-55099 Mainz, Germany\\
$^{40}$ Joint Institute for Nuclear Research, 141980 Dubna, Moscow region, Russia\\
% ...
$^{50}$ Peking University, Beijing 100871, People's Republic of China\\
% ...
$^{62}$ Southeast University, Nanjing 211100, People's Republic of China\\
% ...
$^{64}$ State Key Laboratory of Particle Detection and Electronics, Beijing 100049, Hefei 230026, People's Republic of China\\
% ...
$^{77}$ University of Science and Technology of China, Hefei 230026, People's Republic of China\\
% ...
$^{80}$ University of Turin and INFN, (A)University of Turin, I-10125, Turin, Italy; (B)University of Eastern Piedmont, I-15121, Alessandria, Italy; (C)INFN, I-10125, Turin, Italy\\
$^{81}$ Uppsala University, Box 516, SE-75120 Uppsala, Sweden\\
% ...
$^{86}$ Zhengzhou University, Zhengzhou 450001, People's Republic of China\\

\vspace{0.2cm}
$^{\dagger}$ Deceased\\
% ...
$^{b}$ Also at the Novosibirsk State University, Novosibirsk, 630090, Russia\\
% ...
$^{g}$ Also at State Key Laboratory of Nuclear Physics and Technology, Peking University, Beijing 100871, People's Republic of China\\
% ...

}

\end{document}